\documentstyle[epsfig]{aipproc}
\begin{document}
\title{QED Effective Actions in Inhomogeneous
Backgrounds:\\ Summing the Derivative Expansion}

\author{Gerald V. Dunne}
\address{Department of Physics, University of Connecticut,
Storrs, CT 06269-3046, USA}

\maketitle

\begin{abstract}
The QED effective action encodes nonlinear interactions due to quantum
vacuum polarization effects. While much is known for the special case
of electrons in a constant electromagnetic field (the Euler-Heisenberg
case), much less is known for inhomogeneous backgrounds. Such backgrounds
are more relevant to experimental situations. One way to
treat inhomogeneous backgrounds is the "derivative expansion", in which
one formally expands around the soluble constant-field case. In this talk
I use some recent exactly soluble inhomogeneous backgrounds to perform
precision tests on the derivative expansion, to learn in what sense it
converges or diverges. A closely related question is to find the
exponential correction to Schwinger's pair-production formula for a
constant electric field, when the electric background is inhomogeneous.
\end{abstract}


This talk is concerned with the one-loop QED effective action
\cite{dr,greiner}:
\begin{eqnarray}
S[A]=-\frac{i}{2} \log \det \left( D\hskip -8pt / ~^2+m^2\right)
\label{eff}
\end{eqnarray}
Here $D_\mu=\partial_\mu+i e A_\mu$ is the covariant derivative, and so
$S[A]$ is a functional of the classical background field
$A_\mu(x)$. The effective action is the generating functional for
one-fermion-loop Green's functions, which describe the nonlinear QED
effects to this order. Ideally, one would like to know $S[A]$ for
{\it any} field $A_\mu(x)$, but this is not feasible. However, for the
special case where the field strength tensor, $F_{\mu\nu}=\partial_\mu
A_\nu-\partial_\nu A_\mu$, is uniform, the effective action can be
evaluated in closed form \cite{euler,weisskopf,schwinger}:
\begin{eqnarray}
S=\frac{e^2}{2}(\vec{E}^2-\vec{B}^2) +\frac{2\alpha^2}{45 m^4} \int d^4x\,
\left[(\vec{E}^2-\vec{B}^2)^2+7(\vec{E}\cdot\vec{B})^2\right]+\dots
\label{eh}
\end{eqnarray}
Here the first term is the familiar classical Maxwell action, while the
next term gives the leading quantum correction, which is quartic
in the field strengths. The fine structure constant
$\alpha=\frac{e^2}{4\pi}$ in these units. All higher terms in the
expansion (\ref{eh}) are known explicitly
\cite{euler,weisskopf,schwinger}. It can also be shown that when the
background field is a constant electric field of strength
$E$, the effective action has an exponentially small imaginary part
\begin{eqnarray}
{\rm Im} S = \frac{e^2 E^2}{8\pi^3}\sum_{n=1}^\infty \frac{1}{n^2}
\exp\left[-\frac{m^2\pi n}{eE}\right]
\label{imag}
\end{eqnarray}
This imaginary part of $S$ gives the pair-production rate for
electron-positron production from vacuum in a strong uniform electric
field \cite{euler,schwinger,sauter1}. Note also that it is manifestly
nonperturbative in the coupling constant $e$, in contrast to the
perturbative series expansion in (\ref{eh}).

For a general inhomogeneous background field we need some sort of
approximation to determine the effective action. One such approximation is
the derivative expansion \cite{aitchison,daniel,shovkovy,schubert}, in
which one assumes that $F_{\mu\nu}$ is ``slowly varying'', so that one
can make the following formal expansion
\begin{eqnarray}
S=S^{(0)}[F]+S^{(2)}[F,(\partial F)^2] +\dots 
\label{derivative}
\end{eqnarray}
where $S^{(2n)}$ contains $2n$ derivatives of $F_{\mu\nu}$.  Clearly,
this is a rather formal and ill-defined expansion. In this talk, I
ask two specific questions about this expansion:

$1.$ Does the derivative expansion converge?

$2.$ How is the nonperturbative imaginary piece in (\ref{imag}) modified?

\noindent{These two questions are in fact intimately related.}

First, let us study the convergence properties of the constant-field
Euler-Heisenberg effective action (\ref{eh}). When the
uniform background is purely magnetic, the entire series (we
ignore the Maxwell term) in (\ref{eh}) is
\begin{eqnarray}
S=-\frac{2m^4}{\pi^2}\left(\frac{eB}{m^2}\right)^4\sum_{n=0}^\infty
{2^{2n}{\cal B}_{2n+4}\over (2n+4)(2n+3)(2n+2)}
\left(\frac{eB}{m^2}\right)^{2n}
\label{ehmag}
\end{eqnarray}
This is a "low energy" effective action in the sense that we assume
that the characteristic cyclotron energy of the magnetic background,
$\frac{eB}{m}$, is much less than the electron rest energy, $m$. 
The expansion coefficients in (\ref{ehmag}) involve Bernoulli numbers
${\cal B}_{2n}$, the first few of which are : $\frac{1}{6}$,
$-\frac{1}{30}$, $\frac{1}{42}$, $-\frac{1}{30}$, $\frac{5}{66}$,
$\dots$. The Bernoulli numbers \cite{bernoulli} alternate in sign and grow
factorially in magnitude. In fact, for large $n$,
\begin{eqnarray}
{-2^{2n}{\cal B}_{2n+4}\over (2n+4)(2n+3)(2n+2)}\sim
(-1)^n\frac{1}{8\pi^4}\frac{\Gamma(2n+2)}{\pi^{2n}}\left(1+
\frac{1}{2^{2n+4}}+\frac{1}{3^{2n+4}}+\dots\right)
\label{asymptotic}
\end{eqnarray}
Thus, the Euler-Heisenberg effective action is itself a divergent series!
This is not a bad thing; perturbation theory is typically divergent.  In
many examples (the anharmonic oscillator, the Stark effect, $\phi^3$,
$\phi^4$, QED, ...) perturbation theory is known to be divergent
\cite{leguillou}. Typically, the expansion coefficients diverge as
\begin{eqnarray}
|a_n |\sim  \beta^n\, \Gamma(\gamma\, n +\delta) \left[1+
...\right]
\label{growth}
\end{eqnarray}
at large orders $n$, and usually the parameter $\gamma=1$ or $2$.

Even though a series may be divergent, we can still extract important
and meaningful information from it. One such approach is known as Borel
summation \cite{hardy,carl}.  To motivate this approach, consider the
following expression for the (clearly divergent) alternating series
\begin{eqnarray}
f(g)= \sum_{n=0}^\infty(-1)^n\, n!\, g^{n} &\sim& \frac{1}{g}\int_0^\infty
ds\, {e^{-s/g}\over 1+s}
\label{borel}
\end{eqnarray}
Read from right to left, (\ref{borel}) says that the asymptotic expansion
for small $g$ of the integral on the RHS (which converges for all $g$
positive)  yields the series on the LHS. Read from left to right,
(\ref{borel}) involves a formal interchange of summation and integration
(a step whose validity must be examined carefully in a given case). We
call the integral on the RHS the Borel sum of the divergent series on the
LHS. If we were to try this for a non-alternating series, we might write
analogously
\begin{eqnarray}
f(-g)= \sum_{n=0}^\infty  n!\, g^{n} &\sim& \frac{1}{g}\int_0^\infty ds\,
{e^{-s/g}\over 1-s}
\end{eqnarray}
This is problematic, as there is a pole at $s=1$ on the integration
contour. This pole must be resolved, and in doing so, an
imaginary part arises. Roughly speaking, this imaginary part is of
the form $Im f(-g) = \frac{\pi}{g}\,\exp[-\frac{1}{g}]$, which is
non-perturbative in $g$. Similar arguments apply if the expansion
coefficients grow like
$|a_n|\sim \beta^n\,\Gamma(\gamma\, n +\delta)$, rather than just $n!$ as
in the above example. Then
$f(g)=\sum_n a_n g^n$ behaves like
\begin{eqnarray}
f(g)\sim \frac{1}{\gamma} \int_0^\infty \frac{ds}{s}
\left(\frac{1}{1+s}\right) \left(\frac{s}{\beta g}\right)^{\delta/\gamma}
\, \exp [-\left(\frac{s}{\beta g}\right)^{1/\gamma}]
\label{borelsum}
\end{eqnarray}
\begin{eqnarray}
Im\,f(-g)\sim \frac{\pi}{\gamma} \left(\frac{1}{\beta
g}\right)^{\delta/\gamma} \, \exp
[-\left(\frac{1}{\beta g}\right)^{1/\gamma}]
\label{borelimag}
\end{eqnarray}
I caution that these Borel representations are somewhat formal, as we are
assuming that there are no additional poles or cuts in the complex $g$
plane to prevent the straightforward analytic continuation on which these
expressions are based \cite{hardy,carl}. In this talk our attitude is
exploratory; we shall use these formulae cautiously, with cross-checks,
but defer questions such as uniqueness to more rigorous studies.

It is an instructive exercise to apply this Borel technology to the
alternating divergent Euler-Heisenberg effective action (\ref{ehmag})
\cite{dh3,zhitnitsky,2leh}. Combining the asymptotic form
(\ref{asymptotic}) of the coefficients with the Borel formula
(\ref{borelsum}) we obtain
\begin{eqnarray} 
S= -{e^2B^2\over 8\pi^2}\int_0^\infty {ds\over s^2}
\left(\coth s-{1\over s}-{s\over 3} \right) \,\exp[-{m^2 s\over eB}]
\label{propertime}
\end{eqnarray}
where we have used: $\coth s-1/s=2s/\pi^2\sum_{k=1}^\infty
1/(k^2+s^2/\pi^2)$, and $\sum_{k=1}^\infty1/k^2=\pi^2/6$. But this is
precisely Schwinger's proper-time integral representation
\cite{schwinger,euler,weisskopf} of the effective action, which we see can
be viewed as the Borel sum of the (divergent) Euler-Heisenberg
perturbation series (\ref{ehmag}). If the constant background field is
electric instead of magnetic, the only change perturbatively is to replace
$B^2$ in (\ref{ehmag}) by $-E^2$ (because the Lorentz invariant
combination is $E^2-B^2$). Thus the alternating divergent series
(\ref{ehmag}) becomes a non-alternating divergent series. Applying the
Borel formula (\ref{borelimag}), together with the asymptotic growth rate
in (\ref{asymptotic}), we find an imaginary part in complete
agreement with Schwinger's result (\ref{imag}). So, in this case, our use
of the Borel formulae (\ref{borelsum}-\ref{borelimag}) is
consistent. Further, we see that the Euler-Heisenberg series (\ref{ehmag})
{\it had to be} divergent. If it were not, then there would be no
essential difference between the magnetic and electric cases, and we
would not find any imaginary part for the electric case; so we would
miss the genuine physical effect of vacuum instability. This is somewhat
reminiscent of Dyson's physical argument \cite{dyson} that QED
perturbation theory should be non-analytic as a series in
$\alpha=\frac{e^2}{4\pi}$, since $\alpha$ negative is unstable.

Unfortunately, the pair production rate derived from (\ref{imag}) is
extremely small. Indeed, the critical electric field,
$E_c=\frac{m^2 c^3}{e\hbar}\sim 10^{16}\, V/cm$, where the exponent is
of order 1, is well beyond laboratory static fields. However, using
intense laser fields, it has recently become possible to probe this
critical regime \cite{burke}. These involve inhomogeneous fields, and so
it becomes important to ask how the Euler-Heisenberg analysis is modified
by a field inhomogeneity. Thus we return to our two questions concerning
the convergence properties of the derivative expansion (\ref{derivative})
and how this might modify the imaginary part (\ref{imag}). 

A first guess is that one should look at the high orders of the derivative
expansion to see if it is diverging. Unfortunately this is impossible for
two fundamental reasons. First, the derivative expansion is not actually a
series. This is because as one goes to higher orders, there is a rapid
proliferation of new terms that do not have a
counterpart at previous orders \cite{daniel,shovkovy,schubert}. Second, it
is extremely difficult to calculate any of these terms at very high order,
so we could not expect anyway to observe any asymptotic behaviour of the
expansion coefficients. 

However, we can avoid both these problems at once by considering some
special exactly solvable cases \cite{dh3}. In these cases, the background
inhomogeneity can be characterized by a single scale parameter, so that
the derivative expansion becomes a true series, in inverse powers of this
scale parameter. And since these cases are soluble, we have access to
{\it all orders} of the derivative expansion, so that we can probe the
high order behaviour. So, while we sacrifice generality by concentrating
on specific backgrounds, we gain the ability to perform precise analytic
tests of the derivative expansion. 

To be specific, we will use the fact that for the following backgrounds,
the effective action has been computed in closed-form
\cite{dh1,dh2}.
\begin{eqnarray}
\vec{B}(x)=\vec{B}\,{\rm sech}^2({x\over \lambda})\qquad
or\qquad
\vec{E}(t)=\vec{E}\,{\rm sech}^2({t\over \tau})
\label{ted}
\end{eqnarray}
It has of course been known for many years that the corresponding Dirac
equations are soluble \cite{sauter,naro,baha}. But it is still
non-trivial to perform explicitly the necessary traces so as to
express the effective action (\ref{eff}) as a series or as an integral
representation, in terms of a single integral, just as in the
Euler-Heisenberg case (\ref{ehmag}) or (\ref{propertime}). For the
inhomogeneous magnetic background, $\vec{B}(x)=\vec{B}\,{\rm
sech}^2({x\over \lambda})$, in (\ref{ted}), the exact effective action is
\cite{dh1}
\begin{eqnarray}
S &=& -\frac{m^4}{8\pi^{3/2}} \sum_{j=0}^\infty \sum_{k=1}^\infty 
\frac{1}{(m\lambda)^{2j}}
\left(\frac{2eB}{m^2}\right)^{2k}{\Gamma(2k +j)\Gamma(2k+ j -2){\cal
B}_{2k+2j} \over j!(2k)!\Gamma(2k+j+\frac{1}{2})}
\label{doublesum}
\end{eqnarray}


Several comments are in order concerning this result. First, note that
the series expansion in (\ref{doublesum}) is a {\it double sum}, with 
derivative expansion parameter, $\frac{1}{m\lambda}$, and
perturbative expansion parameter, $\frac{eB}{m^2}$. Second, the expansion
coefficients are known to all orders, and are relatively simple
numbers, just involving the Bernoulli numbers and factorial
factors. 
Third, it has been checked in \cite{dh1} that the first few terms of this
derivative expansion are in agreement with explicit field theoretic
calculations, specialized to this particular background. Fourth, a simple
integral representation for (\ref{doublesum}) can be found in
\cite{dh1}.

Given the explicit series representation in (\ref{doublesum}), we can
check that the series is divergent, but Borel summable (in the sense of
our earlier discussion), in the magnetic case. This can be done in several
ways. One can either fix the order $k$ of the perturbative expansion in
(\ref{doublesum}) and show that the remaining sum is Borel summable, or
one can fix the order $j$ of the derivative expansion in
(\ref{doublesum}) and show that the remaining sum is Borel summable. Or,
one can sum explicitly the $k$ sum, for each $j$, as an integral of a
hypergeometric function, and show that for various values of
$\frac{eB}{m^2}$, the remaining derivative expansion is divergent but
Borel summable. These arguments do not prove rigorously that the double
series is Borel summable, but give a strong numerical
indication that this is the case. 

The case of the inhomogeneous electric field, $\vec{E}(t)=\vec{E}\,{\rm
sech}^2({t\over \tau})$, in (\ref{ted}) can also be solved explicitly
\cite{dh2}. A short-cut to the answer is to note that we can simply
make the replacements, $B^2\to -E^2$, and $\lambda^2\to -\tau^2$, in the
magnetic case result (\ref{doublesum}). In particular this has the
consequence that the alternating divergent series of the magnetic case
becomes a non-alternating divergent series, just as was found in the
Euler-Heisenberg constant-field case. For example, fixing the order $j$
of the derivative expansion, the expansion coefficients behave for large
$k$ (with $j$ fixed) as
\begin{eqnarray}  
a_k^{(j)}=(-1)^{j+k}\frac{\Gamma(2k+j)
\Gamma(2k+j+2){\cal B}_{2k+2j+2}} {\Gamma (2k+3)\Gamma
(2k+j+\frac{5}{2})}\sim  2 
{\Gamma(2k+3j-\frac{1}{2})\over (2\pi)^{2j+2k+2}}
\label{dc}
\end{eqnarray}
Note that these coefficients are non-alternating and grow factorially
with $2k$, as in the form of (\ref{growth}). Applying the Borel dispersion
formula (\ref{borelimag}) gives
\begin{eqnarray} 
{\rm Im} S^{(j)}\sim {m^4\over 8\pi^3}
\left(\frac{eE}{m^2}\right)^{5/2}\, \exp\left[-\frac{m^2\pi}{eE}\right]
\, \frac{1}{j!}\, \left({m^4 \pi\over 4 \tau^2 e^3 E^3}\right)^j
\label{ej}
\end{eqnarray} 
Remarkably, this form can be resummed in $j$, yielding  a leading
exponential form
\begin{eqnarray} 
{\rm Im}S\sim {m^4\over 8\pi^3}
\left(\frac{eE}{m^2}\right)^{5/2}\, \exp\left[-\frac{m^2\pi}{eE}
\left\{1-\frac{1}{4} \left(\frac{m}{eE\tau}\right)^2\right\}\right] 
\label{eresum}
\end{eqnarray}
We recognize the first term in the exponent as the familiar Schwinger
exponent from (\ref{imag}), and so the second term may be viewed as the
leading {\it exponential} correction to the constant-field answer
(\ref{imag}). This is what we set out to find, and we see that it arose
through the divergence of the derivative expansion. I stress that this exponential
correction is not accessible from low orders of the derivative expansion,
such as those studied in \cite{daniel,shovkovy,schubert}.

But the situation is even more interesting than this result
(\ref{eresum}) suggests. For example, we could instead have considered
doing the Borel resummation for the $j$ summations, at each fixed $k$.
Then for large $j$, the coefficients behave as \cite{dh3}
\begin{eqnarray}  
a_j^{(k)}=(-1)^{j+k}\frac{\Gamma(j+2k)
\Gamma(j+2k-2){\cal B}_{2k+2j}} {\Gamma (j+1)\Gamma
(j+2k+\frac{1}{2})}\sim  2^{9/2-2k} 
{\Gamma(2j+4k-\frac{5}{2})\over (2\pi)^{2j+2k}}
\label{dce}
\end{eqnarray}
which once again are non-alternating and factorially growing. Applying the Borel dispersion
formula (\ref{borelimag}) gives
\begin{eqnarray} 
{\rm Im}S^{(k)}\sim {m^{3/2}\over 4\pi^3 \tau^{5/2}}\, {(2\pi
eE\tau^2)^{2k}
\over (2k)!} \, e^{-2\pi m\tau} 
\label{kp}
\end{eqnarray} 
Once again we see that this leading form can be resummed, leading to
\begin{eqnarray} 
{\rm Im}S\sim {m^{3/2}\over 8\pi^3 \tau^{5/2}}\, 
\exp\left[-2\pi m\tau\left(1-\frac{eE\tau}{m}\right)\right]
\label{presum}
\end{eqnarray}
But this leading exponential form of the imaginary part is different from
that obtained in (\ref{eresum}), and moreover, it is different 
from the constant-field case (\ref{imag}). So, what is going on? The
answer is that there are two competing leading exponential behaviours
buried in the double sum (\ref{doublesum}), and the question of which one
dominates depends crucially on the relative magnitudes of the two
expansion parameters. These two expansion parameters are the derivative
expansion parameter, $\frac{1}{m\tau}$, and the perturbative expansion
parameter, $\frac{eE}{m^2}$. Another important parameter is their
{\it ratio}, since this sets the scale of the corresponding gauge
field:
\begin{eqnarray}
\frac{A(t)}{m}=\frac{eE\tau\, {\rm tanh}(t/\tau)}{m}
\sim \frac{eE\tau}{m} ={eE/m^2\over 1/(m\tau)}
\label{scales}
\end{eqnarray}
Thus, we can define a ``non-perturbative'' regime, in which
$\frac{eE\tau}{m}\gg 1$. Then $m\tau\gg\frac{m^2}{eE}$, so that the
dominant exponential factor is $\exp[-\frac{m^2}{eE}]\gg\exp[-2\pi
m\tau]$. In this regime, the leading imaginary contribution to the
effective action is given by the expression (\ref{eresum}), and we note
that it is indeed non-perturbative in form, and the correction in the
exponent is in terms of the small parameter $\frac{m}{eE\tau}\ll 1$.
Alternatively, we can define a ``perturbative'' regime,
in which $\frac{eE\tau}{m}\ll 1$. Then $m\tau\ll\frac{m^2}{eE}$, so that
the dominant exponential factor is
$\exp[-2\pi m\tau]\gg\exp[-\frac{m^2}{eE}]$. In this regime, the leading imaginary contribution to the
effective action is given by the expression (\ref{presum}), and we note
that this is in fact perturbative in nature (despite its
exponential form).

To understand these two regimes more carefully, we can use the WKB
approach developed by Br\'ezin and Itzykson \cite{brezin}, who studied
the case of a sinusoidal electric field: $ E(t)=E \cos(\omega t)$. This
case is not exactly soluble, but a WKB expression for the imaginary part
of the effective action is:
\begin{eqnarray}
Im S\sim \int d^3k\, \exp[-\pi \Omega]
\label{wkb}
\end{eqnarray}
where $\Omega =\frac{2i}{\pi}\int_{\rm tp}
\sqrt{m^2+k_\perp^2+(k_z-eA_z(t))^2}$. If we apply this same WKB analysis
to the (exactly soluble) case $E(t)=E\,{\rm sech}^2(t/\tau)$ in
(\ref{ted}) we find that $\Omega=\tau(\sqrt{m^2+k_\perp^2+(eE\tau+k_z)^2}+
\sqrt{m^2+k_\perp^2+(eE\tau-k_z)^2}-2eE\tau)$. Moreover, this system is
known to be WKB-exact \cite{comtet,baha}. Then when the momentum integrals
in (\ref{wkb}) are done, we obtain precisely the leading results
(\ref{eresum}) or (\ref{presum}), depending on whether we are in the
``non-perturbative'' $\frac{eE\tau}{m}\gg 1$, or ``perturbative''
$\frac{eE\tau}{m}\ll 1$ regime \cite{dh3}. This serves as a useful
cross-check of our Borel analysis.


To conclude, there exist two inhomogeneous backgrounds (one magnetic and
one electric) for which the effective action (\ref{eff}) can be evaluated
exactly, in closed-form, thereby generalizing the well-known
Euler-Heisenberg constant-field case. In this talk I have used these
cases to probe the convergence properties of the derivative expansion. We
learn that the derivative expansion is divergent. In the magnetic case it
appears to be Borel summable, and in the electric case the direct
application of Borel dispersion relations leads to imaginary parts that
are in complete agreement with an independent WKB analysis. Indeed, we can 
argue \`a la Dyson that the derivative expansion must be divergent,
otherwise there would be no exponential correction to the imaginary part
of $S$ for an inhomogeneous electric field. There are two
competing leading exponentials, and in the non-perturbative regime, where
$\frac{eE\tau}{m}\gg 1$, we find an exponential correction (\ref{eresum})
to Schwinger's formula (\ref{imag}). In the perturbative regime, where
$\frac{eE\tau}{m}\ll 1$, a different exponential factor dominates.


\begin{references}
\bibitem{bernoulli} The Bernoulli numbers ${\cal B}_{2n}$ are defined  by
$x/(e^x-1)=\sum_{n=0}^\infty\, {\cal B}_n x^n/n!$.
\end{references}
\end{document}